\documentstyle[preprint,pre,aps,eqsecnum]{revtex}
\tighten
\begin{document}
\draft
\preprint{Submitted to {\sl Physical Review~E}}

\title{Information-theoretic characterization of quantum chaos}

\author{R\"udiger Schack\cite{MileEnd} and Carlton M. Caves}
\address{Center for Advanced Studies,
Department of Physics and Astronomy,\\
University of New Mexico, Albuquerque, NM~87131--1156}
\date{\today}
\maketitle

\begin{abstract}
Hypersensitivity to perturbation is a criterion for chaos based on the question
of how much information about a perturbing environment is needed to keep the
entropy of a Hamiltonian system from increasing. We demonstrate numerically
that hypersensitivity to perturbation is present in the following quantum
maps: the quantum kicked top, the quantum baker's map, the quantum lazy
baker's map, and the quantum sawtooth and cat maps.
\end{abstract}


\section{Introduction}

Present research on quantum chaos focuses mainly on what Berry
\cite{Berry1987} has called quantum chaology, namely ``the study of
semiclassical, but nonclassical, behavior characteristic of systems whose
classical motion exhibits chaos.'' We pursue a different approach, which seeks
to identify intrinsically chaotic features within quantum dynamics.  A
straightforward application of this approach, which attempts to generalize
the classical notion of sensitivity to initial conditions, fails immediately,
for two reasons: first, there is no
quantum analogue of the classical phase-space trajectories in terms of which
sensitivity to initial conditions is defined and, second, the unitarity of
linear Schr\"odinger evolution precludes sensitivity to initial conditions in
the quantum dynamics of state vectors. A classical Hamiltonian system has,
however, in addition to its description in terms of trajectories governed by
the Hamilton equations of motion, an equivalent description in terms of
Liouville probability densities governed by the Liouville equation. Because it
conserves phase-space volumes, the linear Liouville equation shares with the
Schr\"odinger equation a lack of sensitivity to initial conditions. Again, in
Berry's words \cite{Berry1992}, ``the fact that this equation is linear and yet
preserves the chaos of the trajectories disposes of the argument that the
linearity of the Schr\"odinger equation is responsible for the absence of chaos
in quantum mechanics.'' We have recently established a criterion for classical
chaos in terms of Liouville probability densities
\cite{Caves1993b,Schack1992a,Schack1995b}, which we call {\it hypersensitivity
to perturbation\/} and which has a straightforward quantum analogue
\cite{Caves1993b,Schack1993e,Schack1994b}.

Hypersensitivity to perturbation for a quantum system is defined in purely
quantum terms, not in the semiclassical domain.  Though quantum systems show no
sensitivity to initial conditions, due to unitarity, they do show what one
might call sensitivity to parameters in the Hamiltonian, as has been
demonstrated for the quantum kicked top by Peres
{\cite{Peres1991b,Peres1993a}}.  Our approach to quantum chaos could be viewed
as a generalization of Peres's work: whereas Peres compares the time evolution
of the same initial Hilbert-space vector for two slightly different parameter
values in the kicked-top Hamiltonian, we study the distribution of
Hilbert-space vectors arising from the many possible ways in which the
Hamiltonian time evolution can be perturbed through interaction with an
environment. Our approach, however, is embedded in a broader
information-theoretic framework {\cite{Caves1993b,Schack1995b,Schack1994b}}
that is quite different from the viewpoint expressed in {\cite{Peres1993a}}.

In this paper we present numerical evidence that hypersensitivity to
perturbation is present in a variety of quantum maps.  We make no attempt
to develop a general theory of quantum hypersensitivity to perturbation;
rather, our attitude in this paper is exploratory: by investigating numerically
the perturbed dynamics of several quantum systems, we seek to determine
whether hypersensitivity to perturbation provides a criterion for
distinguishing regular from chaotic dynamics in quantum systems. The paper
is organized as follows. Section~\ref{sechyp} defines hypersensitivity to
perturbation and reviews the main results for classical Hamiltonian
systems. Section~\ref{sectop} introduces the model of the quantum kicked top
interacting with a perturbing environment. Section~\ref{secgen}
explains how hypersensitivity to perturbation can be related to the
distribution of vectors in Hilbert space; Sec.~\ref{secres} contains numerical
results for the quantum kicked top. In Sec.~\ref{secmaps} we present numerical
evidence for hypersensitivity to perturbation in a variety of quantum maps on
the unit square. Section~\ref{conc} concludes with brief discussion.

\section{Hypersensitivity to perturbation}         \label{sechyp}

Hypersensitivity to perturbation, in either classical or quantum mechanics,
is defined in terms of information and entropy. The entropy $H$ of an
isolated physical system (Gibbs entropy for a classical system, von Neumann
entropy for a quantum system) does not change under Hamiltonian time
evolution. If the time evolution of the system is perturbed through
interaction with an incompletely known environment $\cal E$, however,
averaging over the perturbation typically leads to an entropy increase
$\Delta H_{\cal S}$. The increase of the system entropy can be limited
to an amount $\Delta H_{\rm tol}$, the {\it tolerable entropy increase},
by obtaining, from the environment, information about the perturbation.
We denote by $\Delta I_{\rm min}$ the minimum information about the
environment needed, on the average, to keep the system entropy below the
tolerable level $\Delta H_{\rm tol}$. A formal definition of the quantities
$\Delta H_{\cal S}$, $\Delta H_{\rm tol}$, and $\Delta I_{\rm min}$ is
given below.

Entropy and information acquire physical content in the presence of a heat
reservoir at temperature $T$. If all energy in the form of heat is ultimately
exchanged with the heat reservoir, then each bit of entropy, i.e., each bit of
{\it missing information\/} about the system state, reduces by the amount
$k_BT\ln2$ the energy that can be extracted from the system in the form of
useful work. The connection between {\it acquired\/} information and work is
provided by Landauer's principle {\cite{Landauer1961,Landauer1988}}, according
to which not only each bit of missing information, but also each bit of
acquired information, has a free-energy cost of $k_BT\ln2$. This
cost, the {\it Landauer erasure cost}, is payed when the acquired information
is erased. Acquired information can be quantified by algorithmic information
{\cite{Chaitin1987a,Zurek1989a,Zurek1989b,Caves1990c,Schack1995a}}.

We are now in a position to define hypersensitivity to perturbation.  We say
that a system is hypersensitive to perturbation if the information $\Delta
I_{\rm min}$ required to reduce the system entropy from $\Delta H_S$ to $\Delta
H_{\rm tol}$ is large compared to the entropy reduction $\Delta H_S-\Delta
H_{\rm tol}$, i.e.,
\begin{equation}
{\Delta I_{\rm min}\over\Delta H_S-\Delta H_{\rm tol}}\gg1\;.
\label{eqhyp}
\end{equation}
The information $\Delta I_{\rm min}$ purchases a reduction
$\Delta H_S-\Delta H_{\rm tol}$ in system entropy, which is equivalent
to an increase in the useful work that can be extracted from the
system; hypersensitivity to perturbation means that the Landauer erasure
cost of the information is much larger than the increase in available
work.

Hypersensitivity to perturbation means that the inequality~(\ref{eqhyp}) holds
for almost all values of $\Delta H_{\rm tol}$.  The inequality~(\ref{eqhyp})
tends always to hold, however, for sufficiently small values of $\Delta H_{\rm
tol}$.  The reason is that for these small values of $\Delta H_{\rm tol}$, one
is gathering enough information from the perturbing environment to track a
particular system state whose entropy is nearly equal to the initial system
entropy.  In other words, one is essentially tracking a particular realization
of the perturbation among all possible realizations.  Thus, for small values of
$\Delta H_{\rm tol}$, the information $\Delta I_{\rm min}$ becomes a property
of the perturbation; it is the information needed to specify a particular
realization of the perturbation.  The important regime for assessing
hypersensitivity to perturbation is where $\Delta H_{\rm tol}$ is fairly
close to $\Delta H_S$, and it is in this regime that one can hope that
$\Delta I_{\rm min}$ reveals something about the system dynamics, rather
than properties of the perturbation.

In {\cite{Schack1995b}} we have shown that a large class of classically chaotic
systems shows {\it exponential hypersensitivity to perturbation}, which is
characterized by an exponential growth with time of the ratio (\ref{eqhyp})
between information and entropy reduction, for a broad range of values of
$\Delta H_{\rm tol}$ below $\Delta H_S$.  The exponential rate of growth is
determined by the Kolmogorov-Sinai entropy of the classical dynamics.  In
preliminary work {\cite{Schack1993e,Schack1994b}} we have applied the notion of
hypersensitivity to perturbation to quantum systems and, through numerical
simulations, have shown that hypersensitivity to perturbation seems to
distinguish regular from chaotic quantum dynamics.  The present paper contains
a more detailed treatment of hypersensitivity to perturbation, including
numerical evidence for a range of prototypic quantum systems.

To define the quantities $\Delta H_{\cal S}$, $\Delta H_{\rm tol}$, and
$\Delta I_{\rm min}$ rigorously in the quantum case, consider a quantum
system $\cal S$ coupled to an environment $\cal E$. We assume that the
total system ${\cal S}\otimes{\cal E}$ is initially in a product state
\begin{equation}
\hat\rho_{\rm total}(t=0)
   = |\psi_0\rangle\langle\psi_0|\otimes\hat\rho_{\cal E}\;,
\end{equation}
combining a pure system state $|\psi_0\rangle$ with a mixed environment
state $\hat\rho_{\cal E}$. The initial von Neumann entropy $H$ of the
system is zero. After evolving for a time $t$ under the action of the
total Hamiltonian ${\hat H}_{\rm total}$, the total system is described
by a density operator $\hat\rho_{\rm total}(t)$, which is generally not
a product state.  The state $\hat\rho_{\cal S}$ of the system $\cal
S$ alone at time $t$ is then obtained by tracing over the environment,
\begin{equation}
\hat\rho_{\cal S} = {\rm tr}_{\cal E}\Bigl(\hat\rho_{\rm total}(t)\Bigr) \;.
\end{equation}
The increase in system entropy due to tracing over the environment is given by
\begin{equation}
\Delta H_{\cal S} =
-{\rm tr}_{\cal S} \Bigl(\hat\rho_{\cal S} \log_2\hat\rho_{\cal S} \Bigr) \;.
\label{eqdelhs}
\end{equation}

Now suppose a measurement is performed on the environment $\cal E$. The
most general measurement is described by a positive-operator-valued measure
(POVM) {\cite{Peres1993a,Kraus}}.  We assume for convenience that the
measurement is described by a POVM that has a discrete set of possible
outcomes $r$ with probabilities $p_r$. For each measurement outcome $r$,
we denote by $\hat\rho_r$ the system state conditional on the measurement
outcome $r$; the conditional system states obey the relation (for a
justification, see Appendix A)
\begin{equation}
\sum_r p_r \hat\rho_r = \hat\rho_{\cal S} \;.
\label{eqsumrho}
\end{equation}
We further define the system entropy conditional on measurement outcome $r$,
\begin{equation}
\Delta H_r = -{\rm tr}_{\cal S} \Bigl( \hat\rho_r \log_2 \hat\rho_r \Bigr) \;,
\label{eqdelhr}
\end{equation}
the average conditional entropy
\begin{equation}
\Delta H = \sum_r p_r \Delta H_r \;,
\end{equation}
and the average information
\begin{equation}
\Delta I = -\sum_r p_r \log_2 p_r \;.
\label{eqdeli}
\end{equation}
Actually $\Delta I$ is a lower bound to the average algorithmic information
needed to specify the measurement outcome $r$. One of us has shown
{\cite{Schack1995a}}, however, that the minimum average algorithmic
information can be bounded above by $\Delta I+1$, so that $\Delta I$ is an
excellent approximation to the minimum average algorithmic information.

The minimum amount of information about the perturbing environment to
keep the system entropy from increasing by more than the tolerable
amount $\Delta H_{\rm tol}$ can now be defined as
\begin{equation}
\Delta I_{\rm min} = \inf_{\Delta H\leq\Delta H_{\rm tol}} \Delta I \;,
\end{equation}
where the infimum is taken over all environment measurements for which
the average conditional entropy increase $\Delta H$ does not exceed
$\Delta H_{\rm tol}$.  In other words, $\Delta I_{\rm min}$ is the
information about the environment it takes to reduce the entropy
increase of the system from $\Delta H_{\cal S}$ (the increase due
to averaging over the perturbation) down to $\Delta H_{\rm tol}$;
i.e., $\Delta I_{\rm min}$ is the minimum information about the
environment needed to purchase a reduction
$\Delta H_{\cal S}-\Delta H_{\rm tol}$ in system entropy.

\section{Quantum kicked top}                   \label{sectop}

\subsection{Kicked-top Hamiltonian}

The quantum model of the kicked top \cite{Frahm1985,Haake1987} describes a
spin-$J$ particle---i.e., an angular momentum vector $\hbar
\hat{\bf J}=\hbar ({\hat J}_x,{\hat J}_y,{\hat J}_z)$,
where $[{\hat J}_i ,{\hat J}_j ]=i\epsilon_{ijk}{\hat J}_k$---whose dynamics in
$(2J+1)$-dimensional Hilbert space is governed by the Hamiltonian

\begin{eqnarray}
\hat H_{\rm top}(t)=(\hbar p/T){\hat J}_z+(\hbar k/{2J}){\hat J}_x^2
\sum_{n=-\infty}^{+\infty}{\delta(t-nT)}\;.
\label{eqhtop}
\end{eqnarray}
The free precession of the spin around the $z$ axis (first term in the
Hamiltonian) is interrupted periodically by sudden kicks or {\it twists\/} at
times $nT$ with twist parameter $k$ (second term in the Hamiltonian). The
angle of free precession between kicks is given by $p$. In this paper we
use $p=\pi/2$.

We look at the time evolution of an initial Hilbert-space vector
$|\psi_0\rangle$ at discrete times $nT$. After $n$ time steps, the evolved
vector is given by
\begin{eqnarray}
|\psi_n\rangle =\hat T_{\rm top}^n |\psi_0\rangle\;,
\end{eqnarray}
where $\hat T_{\rm top}$ is the unitary Floquet operator
\begin{eqnarray}
\hat T_{\rm top}=e^{-i(k/2J){\hat J}_x^2}e^{-i\pi{\hat J}_z/2}\;.
\label{eqqtop}
\end{eqnarray}
The model has a conserved parity, which for half-integer $J$ takes the form
\begin{equation}
\hat S =-i e^{-i\pi\hat J_z}
\label{eqparity}
\end{equation}
and which permits factorization of the matrix representation of the operator
$\hat T_{\rm top}$ into two blocks. Starting from a state with definite
parity, the whole dynamical evolution occurs in the invariant Hilbert subspace
with the given parity.  For half-integer $J$, the dimension of the even-parity
subspace (eigenspace of $\hat S$ with eigenvalue 1) is $J+{1\over2}$. In this
paper, we work with $J=511.5$ in the 512-dimensional even-parity subspace or
with $J=63.5$ in the 64-dimensional even-parity subspace. We consider only the
projection of the initial vector in the even subspace.  Numerical evidence and
symmetry considerations \cite{Peres1991b} suggest that no additional insight is
gained by including the odd-parity subspace.  In any case, the restricted model
can be regarded as a quantum map in its own right, which can be investigated
independently of the behavior of the complete kicked-top model.

\subsection{Choice of initial vector}  \label{secchoice}

Depending on the initial condition, the classical map corresponding to the
Floquet operator~(\ref{eqqtop}) displays regular as well as chaotic behavior
{\cite{Haake1987}}. Following {\cite{Peres1991b}}, we choose initial
Hilbert-space vectors for the quantum evolution that correspond to classical
initial conditions located in regular and chaotic regions of the classical
dynamics, respectively.

The classical map corresponding to the quantum map~(\ref{eqqtop}) is
obtained by introducing the unit vector $(X,Y,Z) \equiv \hat{\bf J}/J$
and performing the limit $J\to\infty$. One obtains {\cite{Haake1987}}
\begin{eqnarray}
&&X'=-Y \nonumber\;,\\
&&Y'=X\cos k Y +Z\sin k Y\;,\label{eqctop} \\
&&Z'=Z\cos k Y -X\sin k Y\;,\nonumber
\end{eqnarray}
which is an area-preserving map of the unit sphere, i.e., an
area-preserving map on the configuration space of a classical spin with fixed
magnitude. Depending on the value of the twist parameter $k$, this map has
regions of chaotic behavior interspersed with regular regions associated with
elliptic cyclic points. In this paper, we are interested in two cyclic points
of the map for $k=3$.  One is an elliptic fixed point of period 1, located
at \cite{D'Ariano1992}
\begin{equation}
Z=\cos\theta=0.455719\;,\;\;\varphi=3\pi/4\;,
\label{eqtopinireg}
\end{equation}
where we use spherical co\"ordinates $\theta = \cos^{-1}\! Z$ and
$\varphi = \tan^{-1}(Y/X)$. For $k=3$ the elliptic fixed
point~(\ref{eqtopinireg}) is surrounded by an oval-shaped regular
region of regular dynamics, extending about $0.3$ radians in the
$\varphi$-direction and about $0.5$ radians in the $\theta$-direction.
The other cyclic point of interest to us is a hyperbolic fixed point
of period 4, which has a positive Lyapunov exponent. It is located
at {\cite{D'Ariano1992}}
\begin{equation}
Z=\cos\theta=0.615950\;,\;\;\varphi=\pi/4\;,
\label{eqtopinichao}
\end{equation}
in the middle of a chaotic region

We want to choose initial vectors for the quantum evolution that correspond
as closely as possible to the classical directions (\ref{eqtopinireg}) and
(\ref{eqtopinichao}). For this purpose, we use {\it coherent states\/}
$|\theta,\varphi\rangle$ \cite{Radcliffe1971,Atkins1971,Perelomov1986},
defined by the relation
\begin{equation}
{\bf n}\cdot\hat{\bf J}|\theta,\varphi\rangle=J|\theta,\varphi\rangle\;,
\end{equation}
where ${\bf n}$ is the unit vector pointing in the direction given by $\theta$
and $\varphi$. All coherent states can be generated by an appropriate rotation
of the state $|J,M=J\rangle=|\theta=0,\varphi=0\rangle$, where $|J,M\rangle$
($M=-J,\ldots,J$) is the simultaneous eigenstate of $\hat J^2$ and
$\hat J_z$ with eigenvalues $J(J+1)$ and $M$, respectively.

The {\it distance\/} between two normalized vectors $|\psi_1\rangle$ and
$|\psi_2\rangle$ is defined as the Hilbert-space angle
\begin{equation}
s(|\psi_1\rangle,|\psi_2\rangle)
=\cos^{-1}(|\langle\psi_1|\psi_2\rangle|)
\label{eqwoott}
\end{equation}
between the two vectors \cite{Wootters1981}.  Consider two coherent states
$|\theta,\varphi\rangle$ and $|\theta',\varphi'\rangle$. In terms of the angle
$\alpha$ between the directions $(\theta,\varphi)$ and $(\theta',\varphi')$,
the distance between the two coherent states is given by \cite{Perelomov1972}
\begin{equation}
\cos[s(|\theta,\varphi\rangle,|\theta',\varphi'\rangle)]=
|\langle\theta,\varphi|\theta',\varphi'\rangle|=
[\cos(\alpha/2)]^{2J}\simeq\exp(-J\alpha^2/4)\;,
\end{equation}
where the approximation is valid for large $J$. Two coherent states can
therefore be regarded as roughly orthogonal if
$\alpha>2J^{-1/2}$ \cite{Peres1991b}.  The {\it size\/} of the coherent
state $|\theta,\varphi\rangle$ is conveniently defined in terms of the $Q$
function
\begin{equation}
Q_{\theta,\varphi}(\theta',\varphi')\equiv
|\langle\theta',\varphi'|\theta,\varphi\rangle|^2=
[\cos(\alpha/2)]^{4J}\equiv Q(\alpha)\;.
\label{eqq}
\end{equation}
Since $Q(2J^{-1/2})\simeq e^{-2}Q(0)$, the $Q$ function of the coherent state
$|\theta,\varphi\rangle$ is very small outside a region of radius $2J^{-1/2}$
centered at the direction $(\theta,\varphi)$.  For the value $J=511.5$ used in
this paper to study regular behavior, one finds a radius of
$2J^{-1/2}\simeq0.09$ radians, less than the size of the regular region around
the elliptic fixed point (\ref{eqtopinireg}).

\subsection{Interaction with an environment}       \label{secenv}

Suppose now that the kicked top interacts with an environment consisting of
a collection of degenerate two-state systems. For the interaction Hamiltonian
we assume the form
\begin{equation}
\hat H_{\rm int}=\hbar g{\hat J}_z\sum_{n=-\infty}^{+\infty}
(\hat\sigma_3)_n \delta ( t - nT - \epsilon) \;.
\label{eqtopint}
\end{equation}
Here $(\hat\sigma_3)_n$ is a Pauli operator for the $n$th two-state system,
$g$ is the interaction strength, and $\epsilon$ is an infinitesimal time
interval. The kicked top thus interacts with a single environment system
at a time, interacting with the $n$th two-state system just after the $n$th
twist, i.e., just after the $n$th unperturbed time step. The $n$th
perturbed time step includes the $n$th unperturbed time step and the
immediately subsequent coupling to the $n$th two-state system. The
total Hamiltonian for system and environment is
\begin{equation}
\hat H_{\rm total} = \hat H_{\rm top} + \hat H_{\rm int} \;,
\end{equation}
where $\hat H_{\rm top}$ is given by Eq.~(\ref{eqhtop}) with $p=\pi/2$.

As was explained in Sec.~\ref{sechyp}, we take the initial state of the total
system to be of the form $|\psi_0\rangle\langle\psi_0|\otimes\hat\rho_{\cal
E}$, where here the initial state $|\psi_0\rangle$ of the system is a coherent
state centered at one of the fixed points (\ref{eqtopinireg}) and
(\ref{eqtopinichao}) and the initial state of the environment, for the
$n$ two-state systems that are relevant for the first $n$ time steps, is
\begin{equation}
\hat\rho_{\cal E} = \sum_{l_1,\ldots,l_n\in\,\{-1,+1\}} 2^{-n}
   |l_1\rangle\cdots|l_n\rangle\langle l_n|\cdots\langle l_1| \;.
\end{equation}
Each of the environment oscillators is with equal probability in the
``lower'' state $|\mbox{$-1$}\rangle$, where
$\hat\sigma_3|\mbox{$-1$}\rangle=-|\mbox{$-1$}\rangle$,
or in the ``upper'' state $|\mbox{$+1$}\rangle$, where
$\hat\sigma_3|\mbox{$+1$}\rangle=|\mbox{$+1$}\rangle$.

Now suppose the kicked top and the environment evolve together for $n$ time
steps. We want to make a measurement on the environment in order to reduce the
entropy increase of the system.  The form of the interaction Hamiltonian
(\ref{eqtopint}) suggests that the natural basis for a measurement of the
environment state is the simultaneous eigenbasis of the operators
$(\hat\sigma_3)_m$.  If the environment is measured in this basis and is
found to be in the state $|l_1\rangle\cdots|l_n\rangle$ ($l_m\in\{-1,+1\}$,
$m=1,\ldots,n$), then the system state after $n$ perturbed steps is given by
\begin{equation}
|\psi_n\rangle =
  \hat T_{\rm top}(l_n) \cdots \hat T_{\rm top}(l_1)\, |\psi_0\rangle \;,
\end{equation}
where
\begin{equation}
\hat T_{\rm top}(l_m) = e^{-i g l_m \hat J_z}
e^{-i(k/2J){\hat J}_x^2}e^{-i\pi{\hat J}_z/2} =
e^{-i g l_m \hat J_z} \hat T_{\rm top}
\end{equation}
is the unperturbed Floquet operator~(\ref{eqqtop}) followed by an
additional rotation about the $z$ axis by angle $gl_m$. Notice
that $\hat T_{\rm top}(l_m)$ commutes with parity~(\ref{eqparity}) and,
hence, does not couple odd- and even-parity subspaces.

The effect of the coupling to the environment, together with the measurement,
is to produce a binary stochastic perturbation at the end of each time
step.  The top rotates by $+g$ if the two-state system is found in the
upper state and by $-g$ if it is found in the lower state, the two
possibilities occurring with equal probability.  After $n$ steps,
there are $2^n$ different equally probable measurement results, each
leading to a pure system state.  Since there is no simple commutation
relation between the operators $\hat T_{\rm top}(-1)$ and
$\hat T_{\rm top}(+1)$, these $2^n$ system states are all different. Any
measurement that fails to distinguish completely between different
environment states $|l_1\rangle\cdots|l_n\rangle$ leads therefore to a
mixed system state. To keep the system state pure, one has to specify
which of the $2^n$ equally probable alternatives is realized. The minimal
information about the environment needed to keep the system entropy at
the level $\Delta H_{\rm tol}=0$ is therefore $\Delta I_{\rm min}=n$
bits. This means that the information about the environment needed to keep
track of the {\it exact\/} system state is unbounded as the number of steps
increases, an effect that is independent of the exact nature of the system
dynamics.

The value of $\Delta I_{\rm min}$ for $\Delta H_{\rm tol}=0$, therefore, is
not a useful measure of the difficulty in keeping the system entropy small. We
must determine how fast $\Delta I_{\rm min}$ decreases if $\Delta H_{\rm tol}$
is increased. Unfortunately, it is difficult to find an optimal measurement
leading to minimal information for an arbitrary value of $\Delta H_{\rm tol}$.
It is possible, however, to find a nearly optimal measurement based on the
following idea. The interaction with the environment (\ref{eqtopint}) is of
a special form that can be described in terms of a randomly perturbed unitary
evolution operator. Each of the $2^n$ possible measurement results corresponds
to one of $2^n$ different {\it perturbation histories\/} obtained by applying
every possible sequence of perturbed unitary evolution operators
$\hat T_{\rm top}(-1)$ and $\hat T_{\rm top}(+1)$ for $n$ steps. Consider
the list $\cal L$ of $2^n$ vectors generated by all perturbation histories.
In the next section, we show how a nearly optimal measurement can be found
in the class of measurements whose outcomes partition the list $\cal L$ into
groups of vectors.

\section{Distribution of vectors in Hilbert space} \label{secdist}

{}From this section on, we always assume that the interaction with the
environment can be described by a stochastic unitary time evolution.  In the
first part of this section, we explore the relation between hypersensitivity to
perturbation and the distribution of the vectors in Hilbert space resulting
from different realizations of the stochastic evolution. In the second part,
we use this relation to investigate hypersensitivity to perturbation in the
quantum kicked top.

\subsection{General theory}                        \label{secgen}

Given the assumption of a stochastic unitary time evolution, the interaction
with the environment at a given time can be described by a list ${\cal
L}=(|\psi_1\rangle,\ldots,|\psi_N\rangle)$ of $N$ vectors in $D$-dimensional
Hilbert space, together with probabilities $q_1,\ldots,q_N$, each vector in
the list corresponding to a particular realization of the perturbation, which
we call a perturbation history.  Averaging over the perturbing environment
leads to a system density operator
\begin{equation}
\hat\rho_{\cal S}=\sum_{j=1}^N q_j |\psi_j\rangle\langle\psi_j| \;,
\end{equation}
with entropy $\Delta H_{\cal S}$ as given by Eq.~(\ref{eqdelhs}).  Consider the
class of measurements on the environment whose outcomes partition the list
$\cal L$ into $R$ groups, labeled by $r=1,\ldots,R$.  We denote by $N_r$ the
number of vectors in the $r$th group ($\sum_{r=1}^{R}N_r=N$). The $N_r$ vectors
in the $r$th group and their probabilities are denoted by
$|\psi^r_1\rangle,\ldots,|\psi^r_{N_r}\rangle$ and $q^r_1,\ldots,q^r_{N_r}$,
respectively. The measurement outcome $r$, occurring with probability
\begin{equation}
p_r=\sum_{i=1}^{N_r}q^r_i \;,
\label{eqpr}
\end{equation}
indicates that the system state is in the $r$th group. The system state
conditional on the measurement outcome $r$ is described by the density operator
\begin{equation}
\hat\rho_r=p_r^{-1}\sum_{i=1}^{N_r}q_i^r|\psi^r_i\rangle\langle\psi^r_i|\;.
\label{eqrhor}
\end{equation}
Using Eqs.~(\ref{eqpr}) and (\ref{eqrhor}), we define the conditional system
entropy $\Delta H_r$, the average conditional entropy $\Delta H$, and the
average information $\Delta I$ as in Eqs.~(\ref{eqdelhr})--(\ref{eqdeli}).

We now describe nearly optimal measurements, i.e., nearly optimal
groupings, in the sense of the discussion in Sec.~\ref{sechyp}. Given
a tolerable entropy $\Delta H_{\rm tol}$, we want to partition the list
of vectors $\cal L$ into groups so as to minimize the information
$\Delta I$ without violating the condition $\Delta H\leq\Delta H_{\rm tol}$.
To minimize $\Delta I$, it is favorable to make the groups as
large as possible. Furthermore, to reduce the contribution to $\Delta H$
of a group containing a given number of vectors, it is favorable to
choose vectors that are as close together as possible in Hilbert space.
Here the distance between two vectors $|\psi_1\rangle$ and $|\psi_2\rangle$
can be quantified in terms of the Hilbert-space angle~(\ref{eqwoott}).
Consequently, to find a nearly optimal grouping, we choose an arbitrary
{\it resolution angle\/} $\phi$ ($0\leq\phi\leq\pi/2$) and group together
vectors that are less than an angle $\phi$ apart.  More precisely, groups
are formed in the following way.  Starting with the first vector
$|\psi_1\rangle$ in the list $\cal L$, the first group is formed of
$|\psi_1\rangle$ and all vectors in $\cal L$ that are within an
angle $\phi$ of $|\psi_1\rangle$. The same procedure is repeated with
the remaining vectors to form the second group, then the third group,
continuing until no ungrouped vectors are left.  This grouping of vectors
corresponds to a partial averaging over the perturbations. To describe a
vector at resolution level $\phi$ amounts to averaging over those details
of the perturbation that do not change the final vector by more than an
angle $\phi$.

For each resolution angle $\phi$, the grouping procedure described above
defines an average conditional entropy $\Delta H\equiv\Delta H(\phi)$ and an
average information $\Delta I\equiv\Delta I(\phi)$. If we choose, for a given
$\phi$, the tolerable entropy $\Delta H_{\rm tol}=\Delta H(\phi)$, then to a
good approximation, the information $\Delta I_{\rm min}$ is given by $\Delta
I_{\rm min}\simeq\Delta I(\phi)$.  By determining the entropy $\Delta H(\phi)$
and the information $\Delta I(\phi)$ as functions of the resolution angle
$\phi$, there emerges a rather detailed picture of how the vectors are
distributed in Hilbert space.  If $\Delta I(\phi)$ is plotted as a function
of $\Delta H(\phi)$ by eliminating the angle $\phi$, one obtains a good
approximation to the functional relationship between $\Delta I_{\rm min}$
and $\Delta H_{\rm tol}$.

It is easy to see that information and entropy obey the inequalities
\begin{equation}
\Delta I(0)\geq\Delta I(\phi)\geq\Delta I(\pi/2)=0\;,
\label{eqinfoineq}
\end{equation}
\begin{equation}
0=\Delta H(0)\leq\Delta H(\phi)\leq\Delta H(\pi/2)\;.
\label{eqhineq}
\end{equation}
The first of the information inequalities in Eq.~(\ref{eqinfoineq}) follows
from the fact that any group at resolution $\phi$ is the union of groups
at resolution $\phi=0$; in words, the average information needed to
specify a group at resolution $\phi=0$ is equal to the average information
needed to specify a group at resolution $\phi$ {\it plus\/} the average
information needed to specify $\phi=0$ groups within the groups at
resolution $\phi$.  The latter of the entropy inequalities in
Eq.~(\ref{eqhineq}) is a consequence of the concavity of the von Neumann
entropy {\cite{Balian1991,Caves1994a}}. A general theorem about average
density operators {\cite{Balian1991,Caves1994a}} shows that, for all
$\phi$,
\begin{equation}
\Delta I(\phi)+\Delta H(\phi)\geq\Delta H(\pi/2)\;. \label{Balian}
\end{equation}
In general, $\Delta I(\phi)$ is a decreasing function of $\phi$,
whereas $\Delta H(\phi)$ is increasing.  This monotonicity can
sometimes be violated, however, because of discontinuous changes in
the grouping of vectors.

As a further characterization of our list of vectors, we calculate the
distribution $g(\phi)$ of Hilbert-space angles
$\phi=s(|\psi\rangle,|\psi'\rangle) =\cos^{-1}(|\langle\psi|\psi'\rangle|)$
between all pairs of vectors $|\psi\rangle$ and $|\psi'\rangle$. For vectors
distributed randomly in $D$-dimensional Hilbert space, the distribution
function $g(\phi)$ is given by {\cite{Schack1994b}}
\begin{equation}
g(\phi)={d{\cal V}_D(\phi)/d\phi\over{\cal V}_D}=
2(D-1)(\sin\phi)^{2D-3}\cos\phi \;.
\end{equation}
Here ${\cal V}_D(\phi)=(\sin\phi)^{2(D-1)}{\cal V}_D$ is the volume contained
within a sphere of radius $\phi$ in $D$-dimensional Hilbert space, and
${\cal V}_D=\pi^{D-1}/(D-1)!$ is the total volume of Hilbert space.  The
maximum of $g(\phi)$ is located at
$\phi=\cos^{-1}\Bigl(1/\sqrt{2(D-1)}\,\Bigr)$; for large-dimensional Hilbert
spaces, $g(\phi)$ is very strongly peaked near the maximum, which is located
at $\phi\simeq\pi/2-1/\sqrt{2D}$, very near $\pi/2$.

As yet a further characterization of the way the vectors are distributed in
Hilbert space, we want to define a quantity that indicates how many dimensions
of Hilbert space are explored by the vectors in the list $\cal L$.  A possible
measure of the number of explored dimensions is the exponential of the entropy,
$2^{\Delta H_{\cal S}}$.  This quantity is bounded above by $D$, the dimension
of Hilbert space, and gets smaller if the dimensions are occupied with
different weights. For example, if two eigenvalues of $\hat\rho_{\cal S}$ are
close to $1/2$ and all the others are close to zero, then $2^{\Delta H_{\cal
S}}\simeq2$, indicating that the vectors are essentially confined to a
two-dimensional subspace.  Unfortunately, if all the vectors in $\cal L$ are
confined to a small sphere with radius $\phi\ll\pi/2$, $\Delta H_{\cal S}$ is
necessarily small just because all the vectors in the group lie along roughly
the same direction in Hilbert space; this is true even if the orthogonal
components of the vectors are evenly distributed over {\it all\/} the
orthogonal directions in Hilbert space. For example, the density operator
describing a uniform distribution of vectors within a sphere of radius
$\phi\ll\pi/2$ has one dominating eigenvalue close to 1 and $D-1$ eigenvalues
that are all equal and close to zero {\cite{Schack1994b}}. Clearly, in
this case $2^{\Delta H_{\cal S}}$ is not an adequate measure of the number of
dimensions explored. On the other hand, if one disregarded the largest
eigenvalue in this example, then the exponential of the entropy would become
a useful measure of the number of explored dimensions.  We therefore introduce
the {\it spread\/} $\Delta H_2$ as the entropy calculated with the largest
eigenvalue of $\hat\rho_{\cal S}$ omitted. The spread is defined as
\begin{equation}
\Delta H_2=-\sum_{k=2}^{D}{\lambda_k\over1-\lambda_1}
\log_2\!\left({\lambda_k\over1-\lambda_1}\right)\;,
\end{equation}
where $\lambda_1\geq\lambda_2\geq\ldots\geq\lambda_{D}$ are the eigenvalues
of the density operator $\hat\rho_{\cal S}$.  By giving different weight to
dimensions corresponding to different eigenvalues of $\hat\rho_{\cal S}$,
the quantity $\lceil2^{\Delta H_2}\rceil$ turns out to be a good indicator
of the number of Hilbert-space dimensions explored by the vectors in the list
$\cal L$, independent of the size of the region occupied by the vectors
($\lceil x\rceil$ denotes the smallest integer greater than or equal
to $x$). In our analysis of the numerical results, we identify the number
of dimensions explored by the entire set of $N$ vectors with the integer
$n_d\equiv\lceil2^{\Delta H_2}\rceil$.

To investigate if a quantum map shows hypersensitivity to perturbation, we use
the following numerical method. We first compute a list of vectors
corresponding to different perturbation histories. Then, for about 50 values of
the angle $\phi$ ranging from 0 to $\pi/2$, we group the vectors in the nearly
optimal way described above. Finally, for each grouping and thus for each
chosen angle $\phi$, we compute the information $\Delta I(\phi)$ and
the entropy $\Delta H(\phi)$. In addition, we compute the angles between all
pairs of vectors in the list and plot them as a histogram approximating the
distribution function $g(\phi)$.

\subsection{Results for quantum kicked top}    \label{secres}

For our numerical study of the kicked top, we use a twist parameter $k=3$ and
either spin $J=511.5$ or spin $J=63.5$. The calculations are done in the
corresponding 512-dimensional or 64-dimensional even-parity subspace; the
even-parity subspace is the eigenspace of the parity~(\ref{eqparity}) with
eigenvalue~$+1$.  Throughout this section, we consider only two initial states.
The first one, the coherent state $|\theta,\varphi\rangle$ with $\theta$
and $\varphi$ given by Eq.~(\ref{eqtopinireg}), is centered in a regular
region of the classical dynamics; we refer to it as the {\it regular initial
state}. The second one, referred to as the {\it chaotic initial state}, is
the coherent state $|\theta,\varphi\rangle$ with $\theta$ and $\varphi$
given by Eq.~(\ref{eqtopinichao}); the chaotic initial state is centered in a
chaotic region of the classical dynamics.

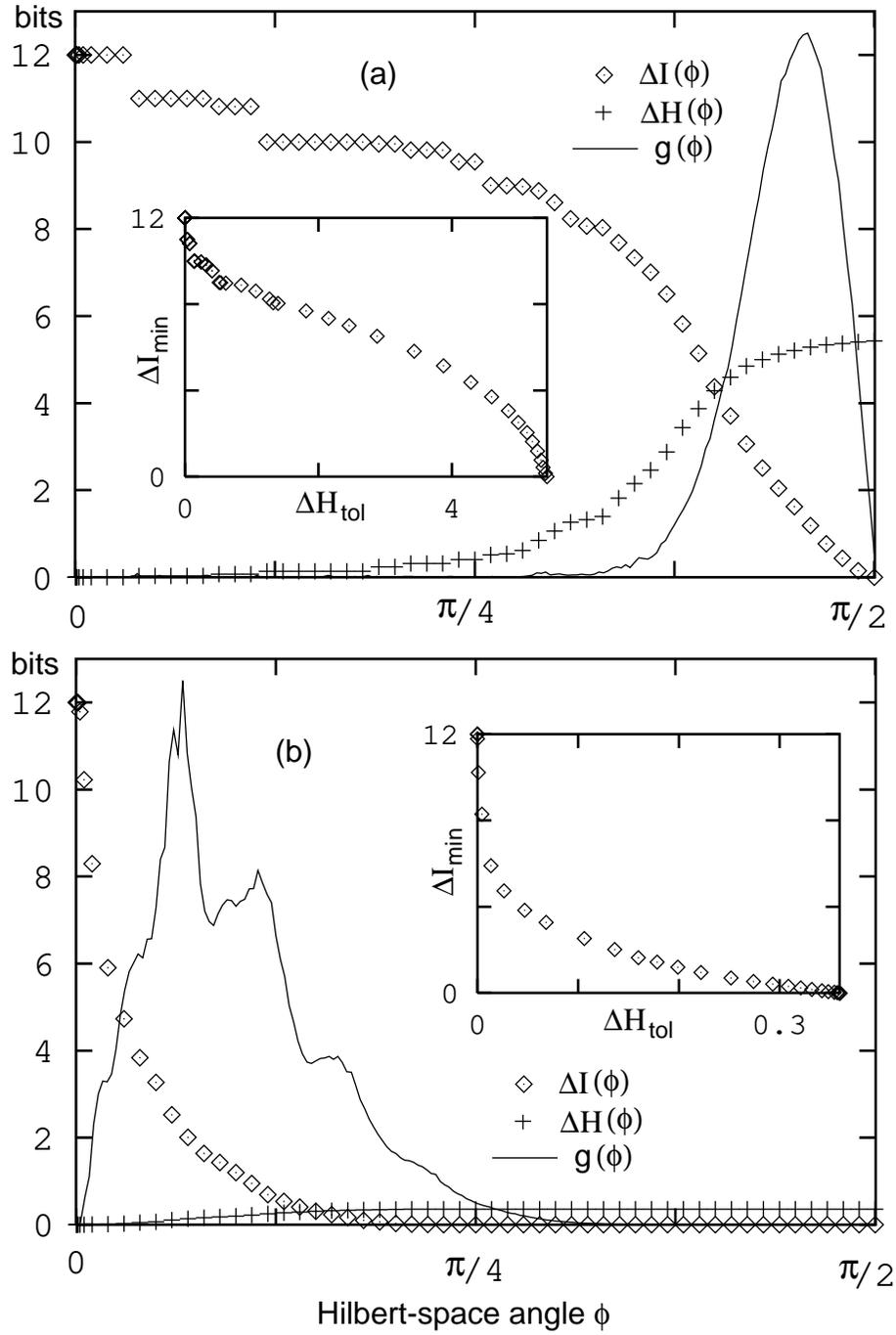
\begin{figure}
\caption{Results characterizing the distribution of Hilbert-space vectors for
the perturbed kicked top with $J=511.5$, $k=3$, and $g=0.003$ after $n=12$ time
steps. The two main diagrams show, as a function of the angle $\phi$, the
distribution $g(\phi)$ of Hilbert-space angles (unnormalized, in arbitrary
units), the average information $\Delta I(\phi)$ to specify a vector at the
resolution given by $\phi$ (in bits), and the average conditional entropy
$\Delta H(\phi)$ (in bits). See the text for a precise definition of these
quantities. In the insets, $\Delta I(\phi)$ is plotted versus $\Delta H(\phi)$
using the same data points as in the main diagrams. The labeling of the inset
axes is motivated by the fact that $\Delta I(\phi)$ is a good approximation to
$\Delta I_{\rm min}$ for $\Delta H_{\rm tol}=\Delta H(\phi)$. (a) Chaotic case,
i.e., initial coherent state $|\psi_0\rangle=|\theta,\varphi\rangle$ centered
in the chaotic region with $\theta$ and $\varphi$ given by
Eq.~(\protect\ref{eqtopinichao}).  Distribution of all $2^{12}$ vectors
generated after 12 perturbed steps.  (b) Regular case, i.e., initial coherent
state centered at the elliptic fixed point given by
Eq.~(\protect\ref{eqtopinireg}). All $2^{12}$ vectors generated after 12
perturbed steps.}
\label{figtop}
\end{figure}

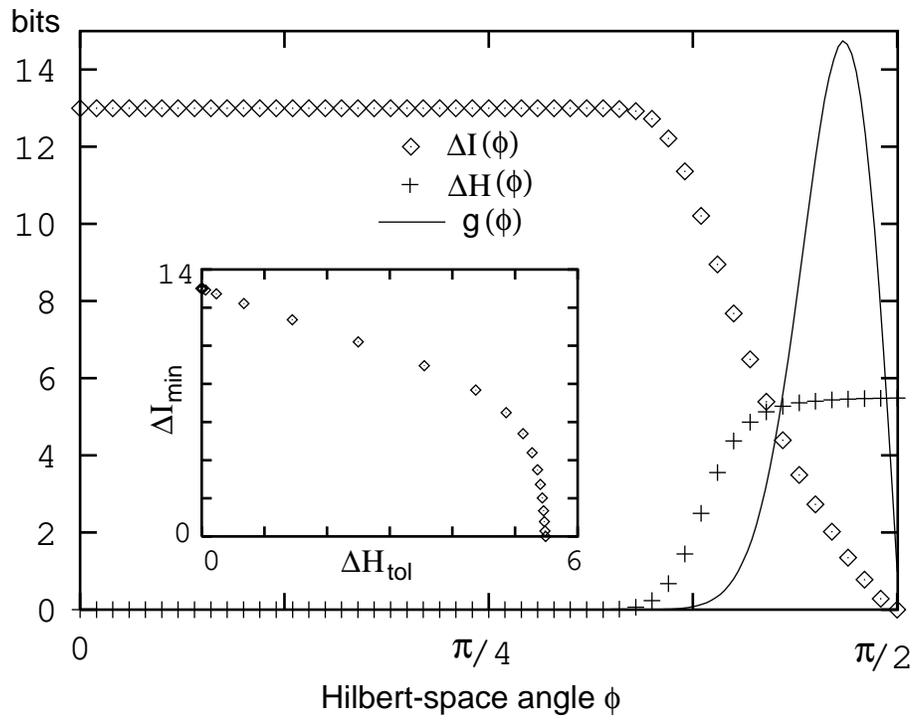
\begin{figure}
\caption{The same quantities as in Fig.~\protect\ref{figtop} plotted using
a list of $2^{13}$ random vectors in 45-dimensional Hilbert space.}
\label{figrandom}
\end{figure}

Figure \ref{figtop} shows results for spin $J=511.5$ and a total number
of $2^{12}=4\,096$ vectors after $n=12$ perturbed steps. The interaction
strength is $g=0.003$, corresponding to perturbing rotations by an angle
$\pm0.003$ radians about the $z$ axis (the perturbation strength in
units of the coherent-state radius $2J^{-1/2}$ is $gJ^{1/2}/2=0.034$).
For Fig.~\ref{figtop}(a), the chaotic initial state was used. The
distribution of Hilbert-space angles, $g(\phi)$, is concentrated at
large angles; i.e., most pairs of vectors are far apart from each other.
The information $\Delta I$ needed to track a perturbed vector at
resolution level $\phi$ is 12 bits at small angles, where each group
contains only one vector. At $\phi\simeq\pi/16$ the information suddenly
drops to 11 bits, which is the information needed to specify one pair of
vectors out of $2^{11}$ pairs, the two vectors in each pair being generated
by perturbation sequences that differ only at the first step.  The sudden
drop of the information to 10 bits at $\phi\simeq\pi/8$ similarly indicates
the existence of $2^{10}$ quartets of vectors, generated by perturbation
sequences differing only in the first two steps.  Figure~\ref{figtop}(a)
suggests that, apart from the organization into pairs and quartets, there
is not much structure in the distribution of vectors for a chaotic initial
state.  The $2^{10}$ quartets seem to be rather uniformly distributed in a
$n_d=\lceil2^{\Delta H_2}\rceil=46$-dimensional Hilbert space, a supposition
supported by a comparison with Fig.~\ref{figrandom}, where results for
$2^{13}=8\,192$ random vectors in 45-dimensional Hilbert space are shown.

The inset in Fig.~\ref{figtop}(a) shows the approximate functional dependence
of the information needed about the perturbation, $\Delta I_{\rm min}$, on the
tolerable entropy $\Delta H_{\rm tol}$, based on the data points $\Delta
I(\phi)$ and $\Delta H(\phi)$. There is an initial sharp drop of the
information, reflecting the grouping of the vectors into pairs and quartets.
Then there is a roughly linear decrease of the information over a wide range
of $\Delta H_{\rm tol}$ values, followed by a final drop with increasing slope
down to zero at the maximum value of the tolerable entropy,
$\Delta H_{\rm tol}=\Delta H_{\cal S}$.  The large slope of the curve near
$\Delta H_{\rm tol}=\Delta H_{\cal S}$ can be regarded as a signature of
hypersensitivity to perturbation. The linear regime at intermediate values
of $\Delta H_{\rm tol}$ is due to the finite size of the sample of vectors:
in this regime the entropy $\Delta H_r$ of the $r$th group is limited by
$\log_2N_r$, the logarithm of the number of vectors in the group.  As is
discussed in more detail in an upcoming publication {\cite{Barnum1995}},
finite-sample-size effects dominate the behavior of the curve as long as the
number of vectors in a typical group is smaller than the effective dimension
of Hilbert space. One therefore expects reduced finite-size effects for larger
samples and smaller Hilbert-space dimensions.

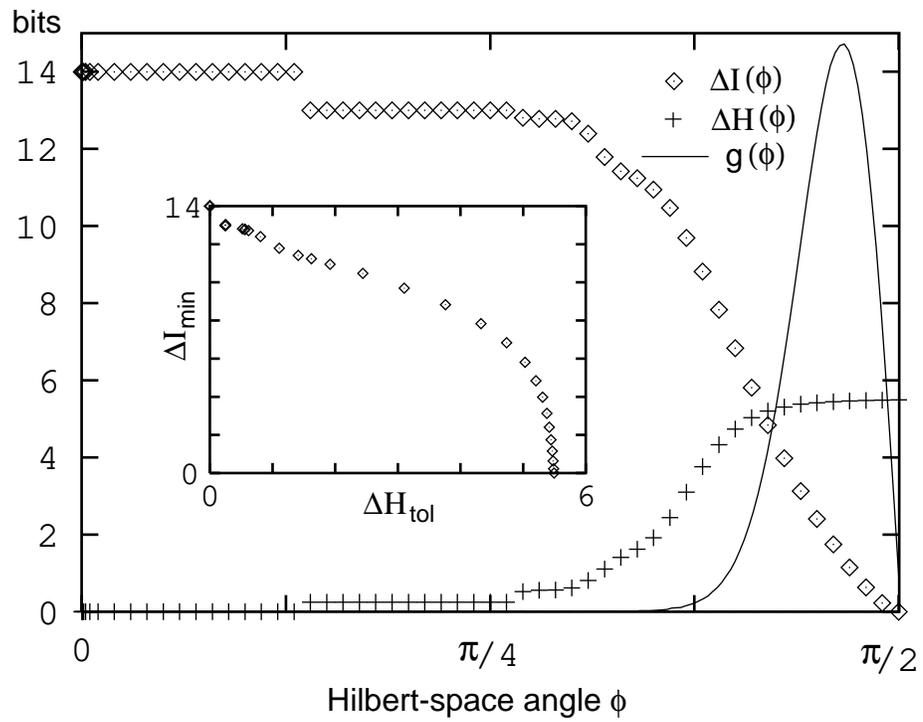
\begin{figure}
\caption{As Fig.~\protect\ref{figtop}(a) (chaotic case), but using $J=63.5$,
$g=0.03$, and all $2^{14}$ vectors generated after 14 perturbed steps.}
\label{figtop64}
\end{figure}

To investigate this prediction, we plot in Fig.~\ref{figtop64} results
for $J=63.5$ and $2^{14}=16\,384$ vectors after 14 perturbed steps,
using an interaction strength $g=0.03$ ($gJ^{1/2}/2=0.12$). As
expected, the inset shows that the large-slope behavior of the
$\Delta I_{\rm min}$ vs.~$\Delta H_{\rm tol}$ curve extends to
higher values of $\Delta I_{\rm min}$ than in the previous case.
Otherwise, Fig.~\ref{figtop64} is very similar to Fig.~\ref{figtop}(a).
Apart from an obvious organization of the vectors into pairs, there
is no apparent structure in the distribution of vectors. The $2^{13}$
pairs of vectors are distributed quasi-randomly in a $n_d=45$ dimensional
Hilbert space, as is confirmed by comparing Fig.~\ref{figtop64} to
the case of $2^{13}$ random vectors in 45 dimensions shown in
Fig.~\ref{figrandom}.

We limit the discussion of the regular initial state to spin $J=511.5$
since, as pointed out in Sec.~\ref{secchoice}, only for a large dimension
of Hilbert space is the regular initial state well localized inside the
regular region around the elliptic fixed point (\ref{eqtopinireg}).
Figure~\ref{figtop}(b) shows data for $2^{12}$ vectors after 12 perturbed
steps in the regular case.  The distribution of perturbed vectors starting
from the regular initial state is completely different from the chaotic
initial condition of Fig.~\ref{figtop}(a).  The angle distribution
$g(\phi)$ is conspicuously non-random: it is concentrated at angles smaller
than roughly $\pi/4$, and there is a regular structure of peaks and valleys.
Accordingly, the information drops rapidly with the angle $\phi$. The number
of explored dimensions is $n_d=2$, which agrees with results of Peres
{\cite{Peres1991b}} that show that the quantum evolution in a regular
region of the kicked top is essentially confined to a 2-dimensional subspace.
The $\Delta I_{\rm min}$ vs.~$\Delta H_{\rm tol}$ curve in the inset bears
little resemblance to the chaotic case. Summarizing, one can say that, in
the regular case, the vectors do not get far apart in Hilbert space,
explore only few dimensions, and do not explore them randomly.

To obtain better numerical evidence for hypersensitivity in the chaotic case
and for the absence of it in the regular case would require much larger
samples of vectors, a possibility that is ruled out by restrictions on
computer memory and time. The hypothesis most strongly supported by our
data is the random character of the distribution of vectors in the chaotic
case. To close this section, we give a crude argument to show that randomness
in the distribution of perturbed vectors implies hypersensitivity to
perturbation.  Consider $N$ vectors distributed randomly in a $D$-dimensional
Hilbert space, where $N\gg D\gg1$, so that $\Delta H_{\cal S}\simeq\log_2 D$.
In this case it was shown in {\cite{Schack1994b}} that
$\Delta I(\phi)\simeq-D\log_2(\sin^2\!\phi)$ and
$\Delta H(\phi)\simeq\sin^2\!\phi\log_2D\simeq\sin^2\!\phi\Delta H_{\cal S}$
if $\Delta I(\phi)<\log_2N$. The information is thus proportional to the
dimension of Hilbert space, $D$, whereas the tolerable entropy is proportional
to $\log_2D$. Eliminating $\phi$ gives
\begin{equation}
\Delta I_{\rm min}\simeq
  D\log_2\!\left({ \log_2D \over \Delta H_{\rm tol} }\right)
  \ge {D\over\ln2}\left(1-{\Delta H_{\rm tol}\over\log_2 D}\right) \simeq
  {D\over\ln D}(\Delta H_{\cal S}-\Delta H_{\rm tol})\gg
  \Delta H_{\cal S}-\Delta H_{\rm tol} \;,
\end{equation}
i.e., hypersensitivity to perturbation.

\section{Quantum maps on the unit square}          \label{secmaps}

\subsection{Quantized unit square}

In this section, we investigate perturbed quantum maps on the unit square.
Our quantization procedure closely follows \cite{Weyl1950} and
\cite{Saraceno1990}. To represent the unit square in $D$-dimensional Hilbert
space, we start with unitary ``displacement'' operators $\hat U$ and $\hat V$,
which produce displacements in the ``momentum'' and ``position'' directions,
respectively, and which obey the commutation relation \cite{Weyl1950}
\begin{equation}
\hat U\hat V = \hat V\hat U\epsilon \;,
\end{equation}
where $\epsilon^D=1$. We choose $\epsilon=e^{2\pi i/D}$.  We further assume
that $\hat V^D=\hat U^D=\eta=\pm1$. The case $\eta=1$ corresponds to periodic
boundary conditions (used in Sec.~\ref{secsaw}), whereas the case $\eta=-1$
corresponds to anti-periodic boundary conditions (used in
Sections~\ref{secbaker} and~\ref{seclazy}).  It follows
\cite{Weyl1950,Saraceno1990} that the operators $\hat U$ and $\hat V$ can
be written as
\begin{equation}
\hat U=e^{2\pi i\hat q}\qquad\mbox{and}\qquad \hat V=e^{-2\pi i\hat p} \;.
\end{equation}
The ``position'' and ``momentum'' operators $\hat q$ and $\hat p$ both have
eigenvalues $(\beta+j)/D$, where the integer $j$ runs over a sequence of $D$
integers and
\begin{equation}
\beta = \left\{  \begin{array}{ll}
0,   & \mbox{if $\eta=1$,} \\
{1\over2}, &\mbox{if $\eta=-1$,} \\
                           \end{array}  \right.
\end{equation}
i.e., $\eta=e^{2\pi i\beta}$.

This quantization procedure fixes the eigenvalues of $\hat q$ and $\hat p$
to lie within some unit interval, but leaves us free to choose the interval.
We use the freedom to write the eigenvalues of $\hat q$ and $\hat p$ as
\begin{equation}
q_j={\beta+j\over D}=p_j\;,\quad j=D_0,\ldots,D_0+D-1\;,
\end{equation}
where the integer $D_0$ can be chosen freely. We denote by $|q_j\rangle$
and $|p_j\rangle$ the eigenvectors of $\hat q$ and $\hat p$ with these
eigenvalues. In Sections~\ref{secbaker} and~\ref{seclazy}, we choose $D_0=0$,
corresponding to $q_j,p_j\in[0,1)$, whereas in Sec.~\ref{secsaw}, we choose
$D_0=-D/2$, corresponding to $q_j,p_j\in[-{1\over2},{1\over2})$.

In the following, we represent Hilbert-space operators by matrices in the
position basis $\{|q_j\rangle\}$.  For consistency of units, let the quantum
scale on phase space be $2\pi\hbar=1/D$. A transformation between the position
basis $\{|q_j\rangle\}$ and the momentum basis $\{|p_j\rangle\}$ is effected
by the operator $G_D$, defined by the matrix elements
\begin{equation}
(G_D)_{kj} = \langle p_k|q_j\rangle =
\sqrt{2\pi\hbar}\; e^{-ip_kq_j/\hbar} =
{1\over\sqrt D}e^{-2\pi i(\beta + k)(\beta + j)/D}\;.
\label{eqfourier}
\end{equation}

Following \cite{Saraceno1990}, we define a family of $D^2$ states
$|q_j,p_k\rangle$ ($j,k=D_0,\ldots,D_0+D-1$), which we call, in analogy to
the continuous case, {\it coherent states}. The fiducial coherent state
$|q_0,p_0\rangle$ (assuming $-D<D_0\leq0$) is defined as the eigenstate with
the smallest eigenvalue of the operator \cite{Saraceno1990}
\begin{equation}
\hat W= 2-{\hat U+\hat U^\dagger\over2}-{\hat V+\hat V^\dagger\over2}
  = 2 - \cos(2\pi\hat q) - \cos(2\pi\hat p) \;.
\end{equation}
The other coherent states are defined by applying the displacement operators
to the fiducial state, i.e.,
\begin{equation}
|q_j,p_k\rangle = e^{-i\pi jk/D}\,\hat U^k \,\hat V^j \, |q_0,p_0\rangle
                =  e^{i\pi jk/D}\,\hat V^j \,\hat U^k \, |q_0,p_0\rangle \;.
\label{eqcoh}
\end{equation}
The coherent states can also be characterized as minimum-uncertainty states
\cite{Saraceno1990}.

The coherent states are an overcomplete set, satisfying
\begin{equation}
{1\over D}\sum_{j,k}|q_j,p_k\rangle\langle q_j,p_k|
=\hat{\openone}=\mbox{(unit operator)}\;.
\end{equation}
Therefore they can be used to define a discrete $Q$ function, representing
an arbitrary state $|\psi\rangle$ via
\begin{equation}
Q_{|\psi\rangle}(q_j,p_k)={1\over D}|\langle q_j,p_k|\psi\rangle|^2\;.
\end{equation}
In terms of the $Q$ function, the coherent state $|q_j,p_k\rangle$ is
represented by roughly circular contours centered at the point $(q_j,p_k)$,
and it occupies an area of about $1/D$ \cite{Saraceno1990}.

For our discussion of the different quantum maps below, we always assume the
initial state $|\psi_0\rangle$ to be a coherent state. The state
$|\psi_n\rangle$ after $n$ unperturbed steps is given by
\begin{equation}
|\psi_n\rangle = \hat T^n\,|\psi_0\rangle \;,
\end{equation}
where $\hat T=\hat T_B$ for the baker's map (Sec.~\ref{secbaker}), $\hat T=\hat
T_L$ for the lazy baker's map (Sec.~\ref{seclazy}), and $\hat T=\hat T_S$ for
the sawtooth and cat maps (Sec.~\ref{secsaw}).

\subsection{Perturbation operator}   \label{secpert}

In Sec.~\ref{secenv}, we presented an example of an interaction with an
environment that can be described as a stochastic perturbation of the unitary
time evolution. Here we assume from the start that at each step a random
perturbation operator is applied following the unperturbed evolution operator
$\hat T$.  The perturbation operator is constructed by analogy to our
previous work \cite{Schack1992a} on the classical baker's map, where the
perturbation consisted in shifting the phase-space pattern by a small amount
in the $p$ direction. The quantum analogue of this perturbation would be an
operator that shifts the $Q$ function of a state by a small amount in the $p$
direction.  Due to the discrete nature of our quantum version of the unit
square, however, a $Q$ function can be shifted rigidly only if it is shifted
by a discrete multiple of $1/D$.

In order to apply a perturbation that does not shift the momentum by a
discrete multiple of $1/D$, one might consider a perturbation operator
$\hat P=e^{-2\pi i\beta\alpha}e^{i\hat q\alpha/\hbar}=
e^{-2\pi i\beta\alpha}e^{2\pi i\hat qD\alpha}$ that ramps up the phase
uniformly in the position basis, i.e.,
$P_{kj}=\delta_{kj}e^{-2\pi i\beta\alpha}e^{2\pi iq_jD\alpha}=
\delta_{kj}e^{2\pi ij\alpha}$.  When $\alpha$ is not a multiple of $1/D$,
however, this perturbation can couple widely separated momentum eigenstates
because of the abrupt phase change between $j=D_0+D-1$ and $j=D_0$. To
avoid this abrupt phase change, we use instead a perturbation operator
$\hat\Pi$ defined in the position basis by
\begin{equation}
\hat\Pi_{kj} =  \left\{  \begin{array}{ll}
\delta_{kj}e^{2\pi ij\alpha}\;,          &\mbox{if $D_0 \leq j<D_0+D/2$,} \\
\delta_{kj}e^{2\pi i(2D_0+D-j-1)\alpha}\;,&\mbox{if $D_0+D/2\leq j<D_0+D$,}
                           \end{array}  \right.
\end{equation}
where $\alpha\ge0$ is the magnitude of the momentum shift ($D\alpha$
is the magnitude of the momentum shift in units of the separation between
momentum eigenstates) and where we restrict attention henceforth to Hilbert
spaces with an even dimension $D$.  Roughly speaking, the operator $\hat\Pi$
shifts the momentum up by $\alpha$ in the left half of the unit square and
down by $\alpha$ in the right half of the unit square.

A perturbed time step consists of first applying the unperturbed
time-evolution operator $\hat T$, followed either by the perturbation
operator $\hat\Pi$ or by its inverse $\hat\Pi^{-1}$, chosen randomly.
After $n$ perturbed time steps, the number of different perturbation
sequences---or histories---is $2^n$, all occurring with equal
probability.

\subsection{Baker's map}                       \label{secbaker}

The classical baker's transformation maps the unit square $0 \leq q,p \leq 1$
onto itself according to
\begin{equation}
(q,p) \longmapsto \left\{  \begin{array}{ll}
\Bigl(2q,{1\over2}p\Bigr)\;,       &\mbox{if $0\leq q\leq{1\over2}$,} \\
\Bigl(2q-1,{1\over2}(p+1)\Bigr)\;, &\mbox{if ${1\over2}<q\leq1$.}
                           \end{array}  \right.
\label{eqcbaker}
\end{equation}
This corresponds to compressing the unit square in the $p$ direction and
stretching it in the $q$ direction, while preserving the area, then cutting it
vertically, and finally stacking the right part on top of the left part---in
analogy to the way a baker kneads dough.

There is no unique way to quantize a classical map. Here we adopt the quantized
baker's map introduced by Balazs and Voros \cite{Balazs1989} and put in more
symmetrical form by Saraceno \cite{Saraceno1990}. We assume anti-periodic
boundary conditions ($\eta=-1$, $\beta={1\over2}$) and $D_0=0$, which leads,
as discussed above, to discrete position and momentum eigenvalues at
half-integer values $q_j=({1\over2}+j)/D=p_j$, $j=0,\ldots,D-1$.  The quantum
baker's map is now defined by the matrix
\begin{equation}
T_B = G_D^{-1} \left( \begin{array}{cc}
                       G_{D/2} & 0 \\ 0 & G_{D/2}
                       \end{array} \right) \;,
\label{eqqbaker}
\end{equation}
where $D$ is even and the matrix elements are to be understood relative to the
position basis.

We simulate the perturbed quantum baker's map for two different
Hilbert-space dimensions, $D=16$ and $D=256$, but using the same
perturbation parameter $\alpha=0.005$ and the same initial coherent state
$|\psi_0\rangle=|1/2D,1/2D\rangle$ (i.e., $j=k=0$) in both cases. In units
of the separation between $\hat p$ eigenstates, this perturbation is 16
times as big for $D=256$ ($D\alpha=1.28$) as for $D=16$ ($D\alpha=0.080$).

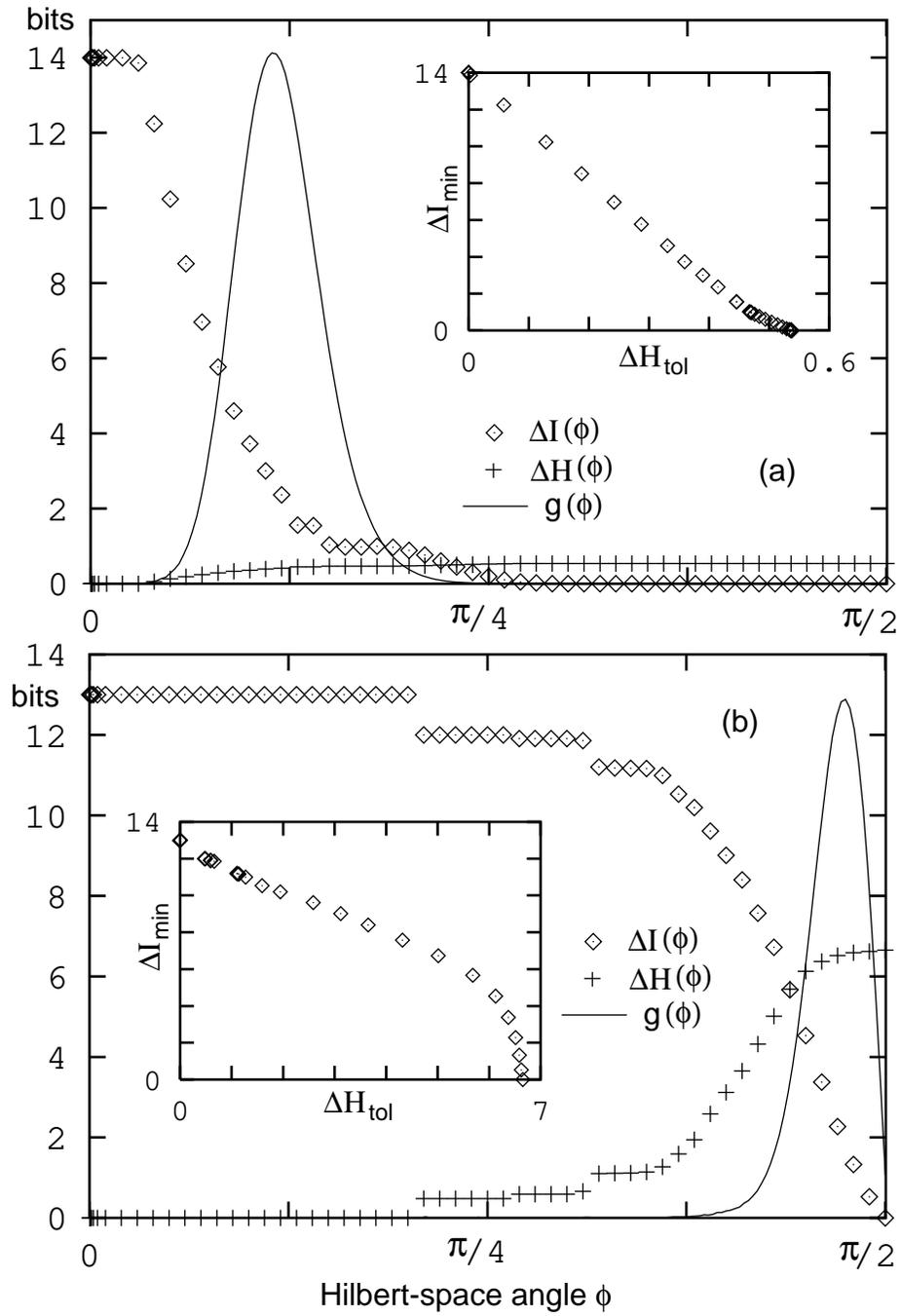
\begin{figure}
\caption{As Fig.~\protect\ref{figtop}, but for the quantum baker's map
with initial coherent state $|\psi_0\rangle=|1/2D,1/2D\rangle$.  (a) $D=16$,
$\alpha=0.005$, all $2^{14}$ vectors after $14$ steps.  (b) $D=256$,
$\alpha=0.005$, all $2^{13}$ vectors after $13$ steps.}
\label{figbaker}
\end{figure}

Figure~\ref{figbaker}(a) shows the results for $D=16$. Because the
perturbation is so small, the vectors after 14 perturbed steps are not
spread very far apart in Hilbert space, a typical angle between vectors
being $\phi=\pi/8$. The information decreases rapidly with resolution angle
$\phi$, and the entropy $\Delta H_{\cal S}$ is less than 1 bit. Nevertheless,
the data have several features characteristic of chaos: (i)~no structure is
apparent in the distribution function $g(\phi)$, indicating a random
distribution of vectors within the small region explored by the vectors;
(ii)~even though the entropy $\Delta H_{\cal S}$ is very small, the number
of explored dimensions is $n_d=6$; (iii)~although the $\Delta I_{\rm min}$
vs.~$\Delta H_{\rm tol}$ curve in the inset is qualitatively different from
the chaotic cases in Figs.~\ref{figtop} and~\ref{figtop64}, it shows
nonetheless that in order to keep the entropy increase $\Delta H_{\rm tol}$
below 10\% of its maximum value $\Delta H_{\cal S}$ requires at least 12
bits of information about the perturbation.  These three features should
be contrasted with the corresponding behavior for the regular initial state
of the quantum kicked top [Fig.~\ref{figtop}(b)].

The effects of the perturbation become more pronounced if the Hilbert-space
dimension is increased to $D=256$, as can be seen in Fig.~\ref{figbaker}(b),
where results for $2^{13}$ vectors after 13 perturbed steps are plotted.  Here,
besides an organization into pairs of vectors a distance of about $3\pi/16$
apart, the distribution of the vectors is similar to the distribution of random
vectors in Fig.~\ref{figrandom}. The vectors explore $n_d=111$ Hilbert-space
dimensions, and the $\Delta I_{\rm min}$ vs.~$\Delta H_{\rm tol}$ curve in
the inset has a steep slope near $\Delta H_{\rm tol}=\Delta H_{\cal S}$, a
signature of hypersensitivity to perturbation.

\subsection{Lazy baker's map}                  \label{seclazy}

The lazy baker's transformation \cite{Lakshminarayan1993a} maps the unit square
$0 \leq q,p \leq 1$ onto itself according to
\begin{equation}
(q,p) \longmapsto \left\{  \begin{array}{ll}
\Bigl(3q,{1\over3}p\Bigr)\;,      &\mbox{if $0 \leq q \leq {1\over3}$,} \\
\Bigl(1-p,q\Bigr)\;,              &\mbox{if ${1\over3} < q \leq {2\over3}$,} \\
\Bigl(3q-2,{1\over3}(p+2)\Bigr)\;, &\mbox{if ${2\over3} < q \leq 1$.}
                           \end{array}  \right.
\label{eqclazy}
\end{equation}
Here the unit square is cut into three vertical stripes, the left and right
stripes are compressed in the $p$ direction and stretched in the $q$ direction
by a factor of 3, the middle stripe is rotated by $90^\circ$, and finally
the three parts are stacked on top of each other---somehow imitating the
action of a lazy baker who neglects to knead the middle portion of the dough.

To define a quantized lazy baker's map, the unit square is quantized in
the same way as for the ordinary baker's map ($\eta=-1$ and $D_0=0$). The
dimension $D$ of Hilbert space is chosen to be a multiple of 6.  The
quantum-mechanical time-evolution operator is {\cite{Lakshminarayan1993a}}
\begin{equation}
T_L = G_D^{-1} \left( \begin{array}{ccc}
                 G_{D/3} & 0 & 0 \\ 0 & I_{D/3} & 0 \\ 0 & 0 & G_{D/3}
                         \end{array} \right) \;,
\label{eqqlazy}
\end{equation}
where $I_{D/3}$ is the identity matrix in $D/3$ dimensions.

The quantum lazy baker's map was introduced {\cite{Lakshminarayan1993a}}
because it allows investigation of regular as well as chaotic evolution, while
being almost as simple as the ordinary baker's map. As we did for the kicked
top, we compare regular and chaotic evolution by choosing different initial
vectors. We use a Hilbert-space dimension $D=216$---even multiples of powers of
3 reflect best the symmetries of the map {\cite{Lakshminarayan1993a}}---and
a perturbation parameter $\alpha=0.005$ ($D\alpha=1.08$) to facilitate
comparison with the results of the last section for the ordinary baker's map.

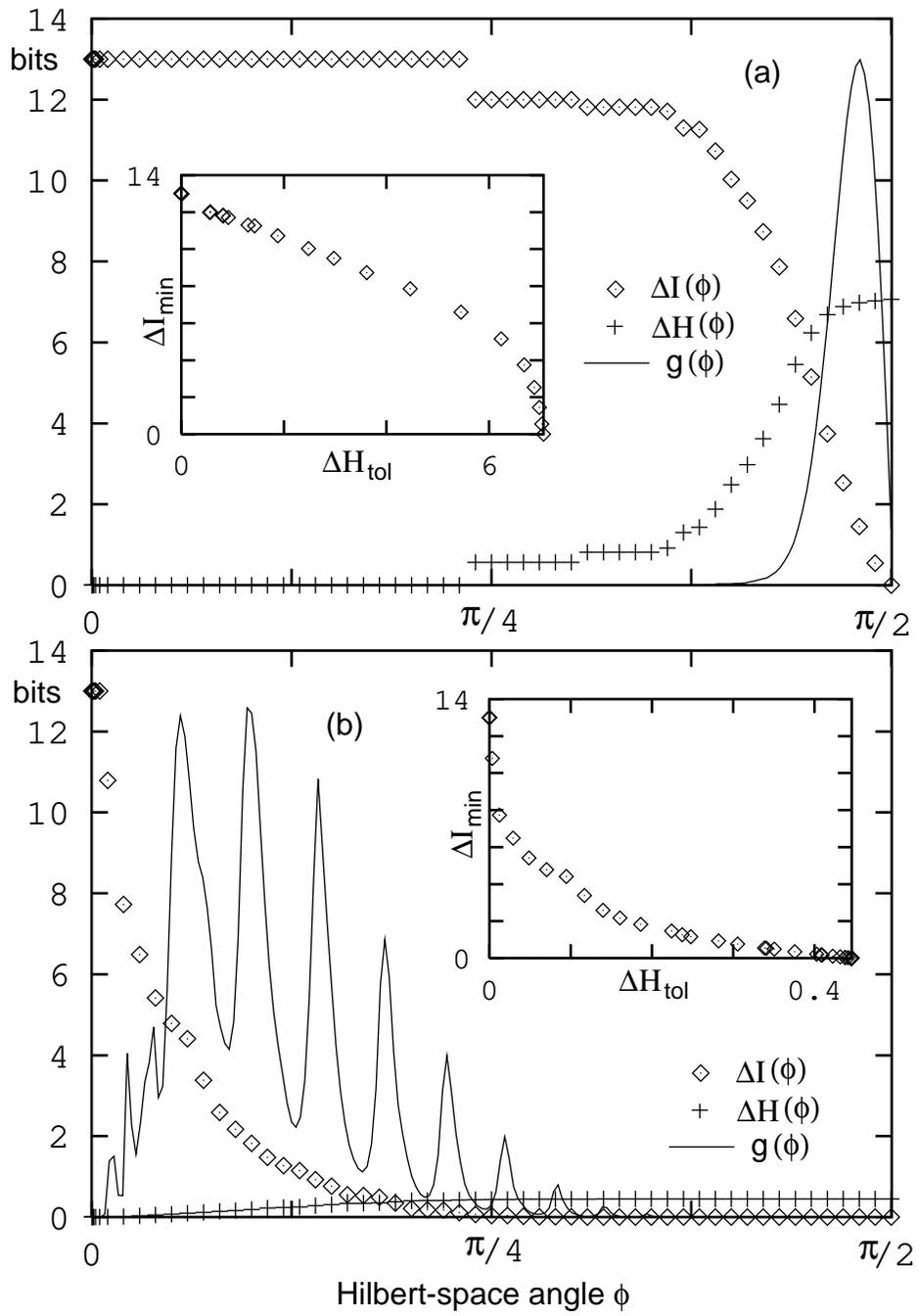
\begin{figure}
\caption{As Fig.~\protect\ref{figtop}, but for the quantum lazy baker's map.
$D=216$,  $\alpha=0.005$, all $2^{13}$ vectors after $13$ steps.
(a) Chaotic case, initial coherent state $|\psi_0\rangle
=|52.5/216,52.5/216\rangle$.  (b) Regular case, initial coherent state
$|\psi_0\rangle=|108.5/216,108.5/216\rangle$.}
\label{figlazy}
\end{figure}

For the chaotic case shown in Fig.~\ref{figlazy}(a), we choose the initial
state to be the coherent state $|\psi_0\rangle=|52.5/216,52.5/216\rangle$
(i.e., $j=k=52$). The $Q$ function of this state, of diameter
$\simeq1/\sqrt{216}\simeq1/15$, covers a region containing a large number
of hyperbolic points separating regular regions in a fractal hierarchy
{\cite{Lakshminarayan1993a}}; thus one expects chaotic behavior for this
initial condition {\cite{Lakshminarayan1993a}}. Indeed, Fig.~\ref{figlazy}(a),
similarly to Fig.~\ref{figbaker}(a), shows that after 13 steps the
$2^{13}$ vectors are distributed quasi-randomly, aside from an organization
into pairs of vectors.  The vectors explore $n_d=136$ dimensions, and the
$\Delta I_{\rm min}$ vs.~$\Delta H_{\rm tol}$ curve has the shape that seems
to be characteristic for chaos.

The results for the regular case, shown in Fig.~\ref{figlazy}(b), are
strikingly different. For this simulation, we choose the initial state
to be the coherent state that is specified by $j=k=108$ (i.e.,
$|\psi_0\rangle=|108.5/216,108.5/216\rangle$). This state, also of diameter
$1/\sqrt{216}\simeq1/15$, is almost entirely confined to the central
square of side length $1/3$, which is simply rotated by $90^\circ$ at
each step.  One expects regular behavior for this trivial time evolution,
and Fig.~\ref{figlazy}(b) confirms this expectation. The peaks and valleys
of the distribution function $g(\phi)$ indicate a regular spacing between
the vectors, the vectors explore only $n_d=2$ dimensions, and the
$\Delta I_{\rm min}$ vs.~$\Delta H_{\rm tol}$ curve has the shape that
we encountered in Fig.~\ref{figtop}(b), which seems to be characteristic
for regular dynamics.

\subsection{Sawtooth and cat maps}             \label{secsaw}

The classical sawtooth map \cite{Percival1987a} on the torus
$q,p\in[-{1\over2},{1\over2})$ is given by
\begin{equation}
(q,p) \longmapsto \Bigl((q+p+Kq)\,\mbox{mod 1},(p+Kq)\,\mbox{mod 1}\Bigr) \;,
\label{eqcsaw}
\end{equation}
where the {\it sawtooth parameter\/} $K$ is an arbitrary real number and
mod 1 is taken relative to the relevant interval $[-{1\over2},{1\over2})$.
For $K=0$, the map is trivial and completely regular, describing a free
rotor. For $K>0$, the map is completely hyperbolic (chaotic). For $K=1$,
it becomes the Arnol'd cat map {\cite{Arnold1968}}.

For the quantized version of the sawtooth maps
\cite{Lakshminarayan1994a,Lakshminarayan1994b}, we assume periodic boundary
conditions ($\eta=1$, $\beta=0$) and $D_0=-D/2$, which leads to discrete
position and momentum eigenvalues at $q_j=p_j=j/D$, $j=-D/2,\ldots,D/2-1$,
where $D$ is even.  The quantum evolution operator is given by the
matrix elements \cite{Lakshminarayan1994a,Lakshminarayan1994b}
\begin{equation}
(T_S)_{jk}=\langle q_j|\hat T_S|q_k\rangle = {e^{-i\pi/4}\over\sqrt{D}}
   \, e^{i\pi Kk^2/D}e^{i\pi(j-k)^2/D} \;.
\end{equation}

For the simulation of the perturbed sawtooth map, we use the Hilbert-space
dimension $D=256$, a very small perturbation parameter $\alpha=0.001$
($D\alpha=0.256$), and the initial coherent state $|\psi_0\rangle=|0,0\rangle$.
We compare the distribution of $2^{13}$ vectors obtained after 13 perturbed
steps for three different values of the sawtooth parameter.

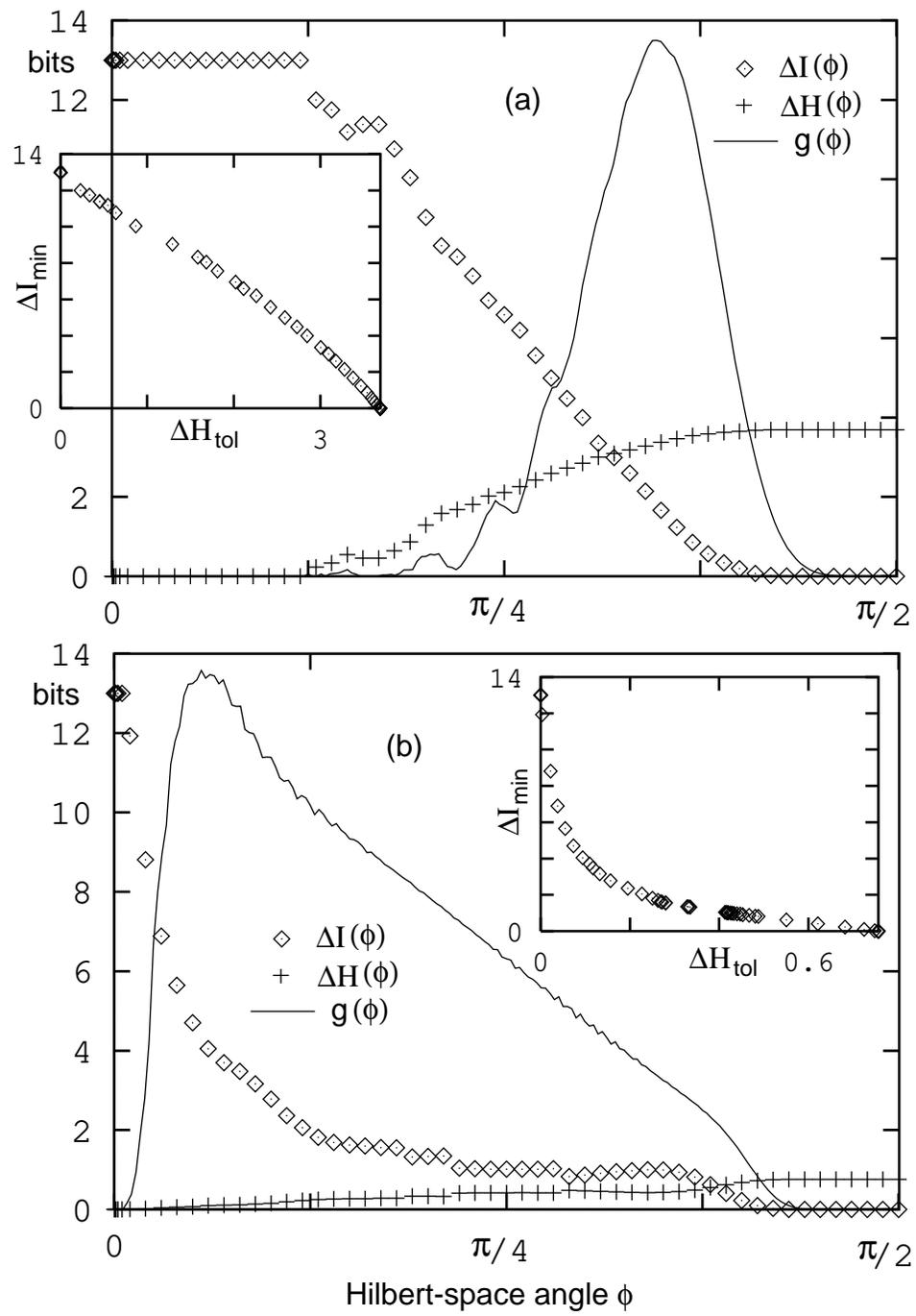
\begin{figure}
\caption{As Fig.~\protect\ref{figtop}, but for the quantum sawtooth maps.
Initial coherent state $|\psi_0\rangle=|0,0\rangle$, $D=256$, $\alpha=0.001$,
all $2^{13}$ vectors after $13$ steps.
(a) Chaotic sawtooth map, $K=0.5$. (b) Regular free rotor, $K=0$.}
\label{figsaw}
\end{figure}

For Fig.~\ref{figsaw}(a) we make a generic choice of the sawtooth parameter,
$K=0.5$, corresponding to chaos in the classical map. As in the behavior
found for the 16-dimensional baker's map in Fig.~\ref{figbaker}{a}, the very
small perturbation strength keeps the vectors from moving very far apart.
Nevertheless, the vectors explore $n_d=38$ dimensions, and to keep the entropy
increase to 10\% of its maximum value requires nearly 12 bits. Altogether,
as for the baker's map in 16 dimensions, we come to the conclusion that the
sawtooth map with $K=0.5$ shows characteristics of hypersensitivity to
perturbation.

In contrast to the generic sawtooth map, consider the free rotor resulting
from a sawtooth parameter $K=0$, a case illustrated in Fig.~\ref{figsaw}(b).
The free rotor shows entirely regular behavior. Although the perturbation
manages to move some vectors far apart from each other, as is evident from
the shape of the distribution function $g(\phi)$, the vectors explore only
$n_d=2$ dimensions of Hilbert space. The $\Delta I_{\rm min}$
vs.~$\Delta H_{\rm tol}$ curve has the shape already found in the regular
cases of the kicked top and the lazy baker's map.

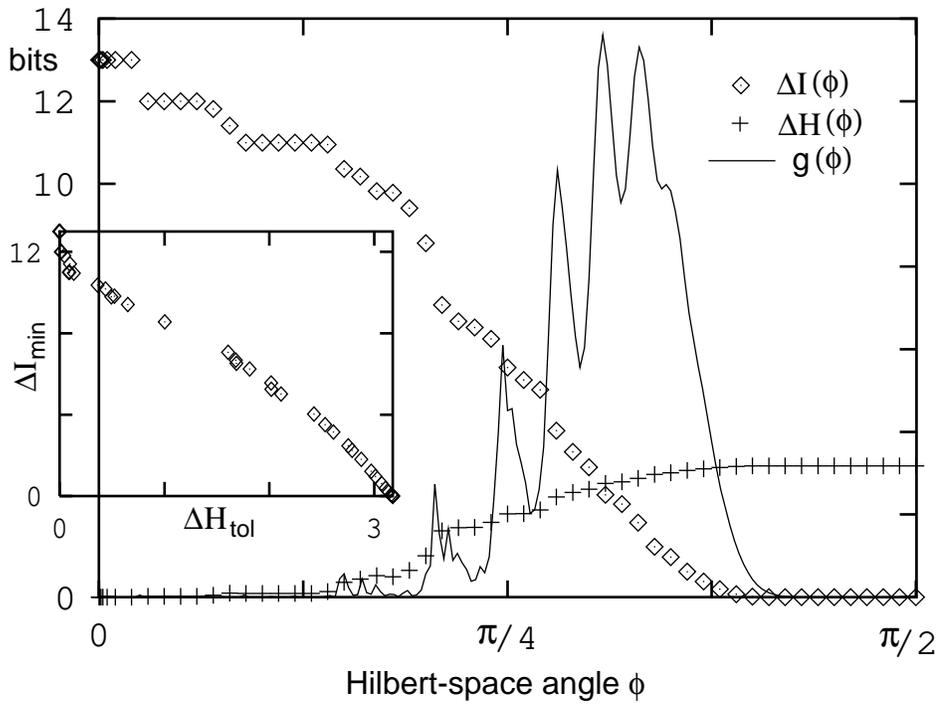
\begin{figure}
\caption{As Fig.~\protect\ref{figtop}, but for the quantum cat map, i.e.,
the quantum sawtooth map with $K=1$.  Initial coherent state
$|\psi_0\rangle=|0,0\rangle$, $D=256$, $\alpha=0.001$, all $2^{13}$ vectors
after $13$ steps.}
\label{figcat}
\end{figure}

Finally, Fig.~\ref{figcat} shows results for the Arnol'd cat map, generated
by a sawtooth parameter $K=1$. Integer values of the sawtooth parameter $K$
are non-generic in the sense that they give the map additional symmetries
{\cite{Percival1987a,Lakshminarayan1994a,Lakshminarayan1994b}}. These
additional symmetries might be responsible for the structure in the angle
distribution $g(\phi)$, apparent in Fig.~\ref{figcat}, which is absent in
the generic case shown in Fig.~\ref{figsaw}(a). Another difference between
the cat map and the generic sawtooth map, which might also be due to the
additional symmetries, is the grouping into pairs, quartets, and possibly
octets, revealed by the sudden drops in the $\Delta I$ vs.~$\phi$ curve
in Fig.~\ref{figcat}, but which is absent in Fig.~\ref{figsaw}(a). As
already discussed for the kicked top in Sec.~\ref{secres}, this grouping
shows up in the $\Delta I_{\rm min}$ vs.~$\Delta H_{\rm tol}$ plot as
an an initial drop of the information.  Apart from this initial drop,
however, the $\Delta I_{\rm min}$ vs.~$\Delta H_{\rm tol}$ curve is
similar to the generic case, indicating that the quantum cat map, like
the generic sawtooth map, is hypersensitive to perturbation.

\section{Conclusion}
\label{conc}

Our aim in this paper is to explore, via numerical simulations, the
notion of hypersensitivity to perturbation in quantum systems, trying
to determine whether hypersensitivity might provide a criterion, as it
does for classical systems, for distinguishing regular from chaotic
quantum dynamics.  What has been learned?  Our simulations indicate that
hypersensitivity is a valuable tool for investigating chaotic quantum
dynamics.  Our key finding is that if a chaotic quantum system is
perturbed stochastically, the different perturbation histories tend
to produce state vectors that are distributed randomly over a large
number of Hilbert-space dimensions, whereas if a regular quantum system
is subjected to a similar perturbation, the resulting state vectors
explore only a few Hilbert-space dimensions and do not explore even
those dimensions randomly.

The nearly random distribution of vectors translates into a
characteristic $\Delta I_{\rm min}$ vs.~$\Delta H_{\rm tol}$ curve,
evident in the insets of Figs.~\ref{figtop}(a), \ref{figrandom},
\ref{figtop64}, \ref{figbaker}(b), and \ref{figlazy}(a), which seems
to be a property of quantum chaos.  This characteristic curve is distinctly
different from the $\Delta I_{\rm min}$ vs.~$\Delta H_{\rm tol}$ curve
produced by regular quantum dynamics [Figs.~\ref{figtop}(b),
\ref{figlazy}(b), and \ref{figsaw}(b)].  After an initial sharp drop
due to perturbation histories that differ only at the first few steps,
the $\Delta I_{\rm min}$ vs.~$\Delta H_{\rm tol}$ curve for chaotic
dynamics has a region of roughly linear decrease that extends over a
broad range of values of $\Delta H_{\rm tol}$; this linear decrease is a
consequence of the finite sample of vectors produced by the perturbation
(always a binary perturbation in our simulations) and thus cannot be
said to be a property of the chaotic dynamics.  The most important
part of the $\Delta I_{\rm min}$ vs.~$\Delta H_{\rm tol}$ curve is
the steep increase in $\Delta I_{\rm min}$ as $\Delta H_{\rm tol}$ is
decreased just below its maximum value of $\Delta H_{\cal S}$.  This
steep increase means the following: to decrease the entropy a small
amount below the entropy that comes from averaging over the stochastic
perturbation requires a large amount of information about the perturbing
environment, much larger than the entropy decrease.  This rapid increase
in information is what we call hypersensitivity to perturbation.  It
seems to be a feature of chaotic quantum dynamics.

These considerations set the stage for further work: to understand in
detail the steep rise in the $\Delta I_{\rm min}$ vs.~$\Delta H_{\rm tol}$
curve and how it is cut off by a finite sample of vectors; to show that
quantum hypersensitivity to perturbation is indeed a property of chaotic
quantum dynamics and not an artifact of particular perturbations; to
study the time dependence of quantum hypersensitivity and perhaps to
extract from the time dependence a quantity like the Kolmogorov-Sinai
entropy of classical chaotic dynamics; and, finally, to develop a general
theory of quantum hypersensitivity to perturbation, which would allow one
to relate it to other properties of chaotic quantum dynamics.

\acknowledgments

The authors profited from discussions with H.~Barnum, D.~Steinbach, and
R.~Menegus. The authors thank the Aspen Center for Physics for making
available its uniquely productive research environment during the period
when this work was first conceived and again when it was being completed.

\appendix

\section{}

Equation~(\ref{eqsumrho}) is true for any measurement on the environment
that is described by a set of {\it completely positive operations\/}
${\cal F}_r$; an operation is a linear, trace-decreasing mapping that
takes positive operators to positive operators and, hence, takes a
density operator to an (unnormalized) post-measurement density operator
\cite{Kraus}.  The joint state of the system and the environment after
a measurement that yields results $r$ is
$$
{{\cal F}_r(\hat\rho_{\rm total})\over p_r}=
{\sum_i\hat A_{ri}\hat\rho_{\rm total}\hat A_{ri}^\dagger\over p_r}\;.
$$
The complete positivity of ${\cal F}_r$ means that it can be written as
the sum involving the operators $\hat A_{ri}$ \cite{Kraus}; since we are
considering a measurement on the environment, these operators are
environment operators.  The probability to obtain measurement result $r$
becomes
$$
p_r={\rm tr}\Bigl({\cal F}_r(\hat\rho_{\rm total})\Bigr)=
{\rm tr}(\hat\rho_{\rm total}\hat E_r)\;,
$$
where
$$
\hat E_r=\sum_i\hat A_{ri}^\dagger\hat A_{ri}
$$
is an element of a POVM.  Completeness of the POVM guarantees that
$$
\sum_r\hat E_r=\hat{\openone}_{\cal E}=\mbox{(environment unit operator).}
$$

The system state after a meaurement that yields result $r$ is obtained
by tracing out the environment,
\begin{equation}
\hat\rho_r=
{{\rm tr}_{\cal E}\Bigl({\cal F}_r(\hat\rho_{\rm total})\Bigr)\over p_r}=
{{\rm tr}_{\cal E}(\hat\rho_{\rm total}\hat E_r)\over p_r}\;.
\label{eqappendix}
\end{equation}
The crucial step here uses the cyclic property of the trace to move the
operators $\hat A_{ri}$ to the right side of the trace; this step is
allowed within an environment trace because these operators are, by
assumption, environment operators.  The important content of
Eq.~(\ref{eqappendix}) is that even though there are many operations
${\cal F}_r$ that are consistent with the POVM $\hat E_r$, the system
state after a measurement on the environment depends only on the POVM.
Equation~(\ref{eqsumrho}) follows from summing over $r$ and using the
completeness of the POVM:
$$
\sum_rp_r\hat\rho_r={\rm tr}_{\cal E}(\hat\rho_{\rm total})
=\hat\rho_{\cal S}\;.
$$

\newpage

\begin{thebibliography}{10}

\bibitem[*]{MileEnd}
Present address: Department of Physics, Queen Mary and Westfield College,
University of London, Mile End Road, London E1~4NS, UK (e-mail:
r.schack@qmw.ac.uk)

\bibitem{Berry1987}
M.~V. Berry, Proc.\ Roy.\ Soc.\ Lond.\ A {\bf 413},  183  (1987).

\bibitem{Berry1992}
M.~V. Berry,  in {\em New Trends in Nuclear Collective Dynamics}, edited by Y.
  Abe, H. Horiuchi, and K. Matsuyanagi (Springer, Berlin, 1992), p.\ 183.

\bibitem{Caves1993b}
C.~M. Caves,  in {\em Physical Origins of Time Asymmetry}, edited by J.~J.
  Halliwell, J. P\'erez-Mercader, and W.~H. Zurek (Cambridge University Press,
  Cambridge, England, 1993), p.\ 47.

\bibitem{Schack1992a}
R. Schack and C.~M. Caves, Phys.\ Rev.\ Lett.\ {\bf 69},  3413  (1992).

\bibitem{Schack1995b}
R. Schack and C.~M. Caves,   submitted to Phys.\ Rev.~E [e-print
chao-dyn/9506002].

\bibitem{Schack1993e}
R. Schack and C.~M. Caves, Phys.\ Rev.\ Lett.\ {\bf 71},  525  (1993).

\bibitem{Schack1994b}
R. Schack, G.~M. D'Ariano, and C.~M. Caves, Phys.\ Rev.\ E {\bf 50},  972
  (1994).

\bibitem{Peres1991b}
A. Peres,  in {\em Quantum Chaos}, edited by H.~A. Cerdeira, R. Ramaswamy,
  M.~C. Gutzwiller, and G. Casati (World Scientific, Singapore, 1991), p.\
  73.

\bibitem{Peres1993a}
A. Peres, {\em Quantum Theory: Concepts and Methods\/} (Kluwer Academic
  Publishers, Dordrecht, The Netherlands, 1993).

\bibitem{Landauer1961}
R. Landauer, IBM J. Res.\ Develop.\ {\bf 5},  183  (1961).

\bibitem{Landauer1988}
R. Landauer, Nature {\bf 355},  779  (1988).

\bibitem{Chaitin1987a}
G.~J. Chaitin, {\em Algorithmic Information Theory\/} (Cambridge University
  Press, Cambridge, England, 1987).

\bibitem{Zurek1989a}
W.~H. Zurek, Nature {\bf 341},  119  (1989).

\bibitem{Zurek1989b}
W.~H. Zurek, Phys.\ Rev.\ A {\bf 40},  4731  (1989).

\bibitem{Caves1990c}
C.~M. Caves,  in {\em Complexity, Entropy, and the Physics of Information},
  edited by W.~H. Zurek (Addison Wesley, Redwood City, California, 1990), p.\
  91.

\bibitem{Schack1995a}
R. Schack,   submitted to Phys.\ Rev.~E [e-print hep-th/9409022].

\bibitem{Kraus}
K.~Kraus, {\em States, Effects, and Operations: Fundamental Notions of
  Quantum Theory\/} (Springer, Berlin, 1983).

\bibitem{Frahm1985}
H. Frahm and H.~J. Mikeska, Z. Phys.\ B {\bf 60},  117  (1985).

\bibitem{Haake1987}
F. Haake, M. Ku\'{s}, and R. Scharf, Z. Phys.\ B {\bf 65},  381  (1987).

\bibitem{D'Ariano1992}
G.~M. D'Ariano, L.~R. Evangelista, and M. Saraceno, Phys.\ Rev.\ A {\bf 45},
  3646  (1992).

\bibitem{Radcliffe1971}
J.~M. Radcliffe, J. Phys.\ A {\bf 4},  313  (1971).

\bibitem{Atkins1971}
P.~W. Atkins and J.~C. Dobson, Proc.\ Roy.\ Soc.\ Lond.\ A {\bf 321},  321
  (1971).

\bibitem{Perelomov1986}
A.~M. Perelomov, {\em Generalized Coherent States\/} (Springer, Berlin, 1986).

\bibitem{Wootters1981}
W.~K. Wootters, Phys.\ Rev.\ D {\bf 23},  357  (1981).

\bibitem{Perelomov1972}
A.~M. Perelomov, Commun.\ Math.\ Phys.\ {\bf 26},  222  (1972).

\bibitem{Balian1991}
R. Balian, {\em From Microphysics to Macrophysics}, Vol.~1 (Springer, Berlin,
  1991).

\bibitem{Caves1994a}
C.~M. Caves and P.~D. Drummond, Rev.\ Mod.\ Phys.\ {\bf 66},  481  (1994).

\bibitem{Barnum1995}
H.~Barnum, R.~Schack, and C.~M. Caves, in preparation.

\bibitem{Weyl1950}
H. Weyl, {\em The Theory of Groups and Quantum Mechanics\/} (Dover, New York,
  1950).

\bibitem{Saraceno1990}
M. Saraceno, Ann.\ Phys.\ {\bf 199},  37  (1990).

\bibitem{Balazs1989}
N.~L. Balazs and A. Voros, Ann.\ Phys.\ {\bf 190},  1  (1989).

\bibitem{Lakshminarayan1993a}
A. Lakshminarayan and N.~L. Balazs, Ann.\ Phys.\ {\bf 226},  350  (1993).

\bibitem{Percival1987a}
I.~C. Percival and F. Vivaldi, Physica D {\bf 27D},  373  (1987).

\bibitem{Arnold1968}
V.~I. Arnold and A. Avez, {\em Ergodic Problems of Classical Mechanics\/}
  (Benjamin, New York, 1968).

\bibitem{Lakshminarayan1994a}
A. Lakshminarayan and N.~L. Balazs, to be published [e-print chao-dyn/9307005].

\bibitem{Lakshminarayan1994b}
A. Lakshminarayan, Phys.\ Lett.\ A {\bf 192},  345  (1994).

\end{thebibliography}
\end{document}